\documentclass[aps,twocolumn,pra,tightenlines,floatfix,showpacs]{revtex4}
\usepackage[dvips]{graphicx}
\usepackage[english]{babel}
\usepackage{amsmath}
\usepackage{amssymb}
\usepackage{bm,times}
\newcommand{\Ek}{E_{\mathbf{k}}}
\newcommand{\uk}{u_{\mathbf{k}}}
\newcommand{\vk}{v_{\mathbf{k}}}

\newcommand{\up} {\uparrow}
\newcommand{\down} {\downarrow}
\newcommand{\vect}[1] {\mathbf{#1}}

\begin{document}

\title{Comparative Study of BCS-BEC Crossover Theories above $T_c$: 
the Nature of the Pseudogap in Ultra-Cold Atomic Fermi Gases}

\author{Chih-Chun Chien, Hao Guo, Yan He, and K. Levin}

\affiliation{James Franck Institute and Department of Physics,
University of Chicago, Chicago, Illinois 60637, USA}

\date{\today}

\begin{abstract}
This paper presents a comparison of two finite-temperature BCS-Bose
Einstein condensation (BEC) crossover theories above the transition
temperature: Nozieres Schmitt-Rink (NSR) theory and finite
$T$-extended BCS-Leggett theory. The comparison is cast in the form of
numerical studies of the behavior of the fermionic spectral function
both theoretically and as constrained by (primarily) radio frequency
(RF) experiments.  Both theories include pair fluctuations and exhibit
pseudogap effects, although the nature of this pseudogap is very
different. The pseudogap in finite $T$-extended BCS-Leggett theory is
found to follow a BCS-like dispersion which, in turn, is associated
with a broadened BCS-like self energy, rather more similar to what is
observed in high temperature superconductors (albeit, for a $d$-wave
case). The fermionic quasi-particle dispersion is different in NSR
theory and the damping is considerably larger. We argue that the two
theories are appropriate in different temperature regimes with the
BCS-Leggett approach more suitable nearer to condensation. There
should, in effect, be little difference at higher $T$ as the pseudogap
becomes weaker and where the simplifying approximations used in the
BCS-Leggett approach break down.  On the basis of momentum-integrated
radio frequency studies of unpolarized gases, it would be difficult to
distinguish which theory is the better.  A full comparison for
polarized gases is not possible since there is claimed to be
inconsistencies in the NSR approach (not found in the BCS-Leggett
scheme).  Future experiments along the lines of momentum resolved
experiments look to be very promising in distinguishing the two
theories.
\end{abstract}


\maketitle

\section{Introduction}
\label{sec:intro}
The behavior of
ultracold superfluid Fermi gases
continues to attract attention from
the experimental and theoretical communities.  
Through a Feshbach resonance
\cite{KetterleVarenna,Chinreview}
which tunes the strength of the attractive
interaction, these trapped gases can exhibit a crossover from BCS to
Bose Einstein condensation (BEC). 
Unlike
the Bose superfluids, at the time of their discovery, the Fermi superfluids
are not associated with any ready-made theory (such as
Gross Pitaevskii, or Bogoliubov theory for bosons).
This provides
an opportunity for theorists to work hand in hand with
experimentalists to arrive at the same level of understanding
of the fermionic as was reached for the bosonic superfluids.
While the Fermi systems are harder to address than the
Bose counterparts, the payoff for progress is great.
Moreover,
there is a general belief that these systems may lead to
important insights into high temperature superconductor (HTSCs), 
in part because the HTSCs exhibit an anomalously short coherence
length which suggests that they may be mid-way between BCS and
BEC
\cite{LeggettNature,Ourreview,Sawatzky,Chienlattice2}.

There is, as yet, no clear consensus about the theory which
underlies BCS-BEC crossover although there are three rather well
studied analytic many body theories which have emerged.
The goal of the present paper is to present a comparison and
assessment of these approaches with a particular focus on the two
of the three, which seem most reliable.
In addition to assessing the theoretical approaches,
we address physical consequences and how the theories may
be differentiated through the behavior of the
centrally important spectral function and related density of
states. We do this in the context of radio-frequency-(RF) based probes.

What is among the most interesting features of BCS-BEC
crossover is the fact that the normal state (out of which 
superfluidity forms) is different from the normal state (Fermi liquid) 
associated with strict BCS theory. The
normal state of, for example, a unitary gas
consists of pre-formed pairs which persist below
$T_c$ in the form of non-condensed pair excitations of the condensate.
This excitation branch is in addition to the usual gapped
fermionic excitations.
The normal state is often said to exhibit a ``pseudogap" which has features
in common with the exotic normal state of the high temperature
superconductors. This pseudogap \cite{Ourreview,Strinaticuprates}
reflects the formation of quasi-bound
pairs which in turn require an energy input (called $\Delta$) in
order to break the pairs and create fermionic excitations.
Physically, what differs from one crossover theory to another 
\cite{firstordertransitionpapers,Strinaticuprates,Ourreview,Drummond3}
is the nature
of these non-condensed or pre-formed pairs which, respectively,
appear below
and above
$T_c$.
Unlike the pair fluctuations of traditional superconductors (which
are associated with low dimensionality and impurity effects) these pairs
are present because of a stronger-than-BCS attractive interaction.
As a consequence, the pairing gap $\Delta$ persists to temperatures
which can be several times $T_c$ for the case of the unitary gases.

In this paper we address the
temperature dependence of the spectral function particularly
in the normal state. 
The density of states (DOS), which
can be obtained from the spectral function, will also be presented. We
compare with experiments in the context of
RF spectra of both unpolarized and polarized
Fermi gases. 
Quantum Monte Carlo simulations
\cite{QMCTc,MCpair} provide useful information such as the superfluid
transition temperature $T_c$, entropy, condensate fraction, etc.
and recently reveal evidence of non-condensed pairs \cite{Bulgacpairs}
along with a pseudogap \cite{BulgacPG} in the normal phase. Our focus
is on two different finite-temperature BCS-BEC crossover theories and
we present a detailed comparison of the results obtained from the two
theories as well as an assessment of other BCS-BEC crossover theories.

\subsubsection{Analysis of Different Crossover Theories}

A fair amount of controversy \cite{Ourreview,CompareReview,Drummond5,HaussmannRF}
has surfaced in the literature regarding the
three alternative analytic pairing fluctuation schemes. In this paper we
address some of these issues and clarify misleading claims.
At this early stage of understanding we do not believe it is suitable
to invoke (possibly fortuitous)
fits to particular experimental or Monte-Carlo derived
numbers to establish which of these theories is ``best". Rather
in line with the goal of this paper,
one has to look at the differences at a more general level.
One has, furthermore, to subject these theories to careful
consistency tests.

Each of the three many body approaches is associated with a different
ground state. Thus far, only one of these can be written down analytically.
In this context we note that one
can trace the historical origin of the BCS-BEC literature to the observation
that the simplest BCS-like wavefunction
\begin{equation}
\Psi_0=\Pi_{\bf k}(\uk+\vk
c_{\vect
k\up}^\dagger
c_{-\vect k\down} ^\dagger)
|0\rangle ,
\label{eq:1}
\end{equation}
is much more general than originally presumed
\cite{Leggett,Eagles,NSR}. To implement this generalization, all
that is required is that one
solve for the
two variational parameters $\uk$ and $\vk$ 
in concert with a self consistent condition on the fermionic
chemical potential $\mu$. As the attraction is increased,
$\mu$ becomes different from the
Fermi energy $E_F$, and in particular, in
the BEC regime, $\mu$ is
negative.
This ground state is often called the ``BCS-Leggett" state
and the two variational parameters $\uk$ and $\vk$ can
be converted to two more physically accessible parameters associated
with the zero temperature gap (or equivalently order parameter)
and $\mu$.

The three theories currently of interest can be related to a t-matrix
scheme.
Within a given t-matrix scheme one treats the fermionic
self energy and the pair-propagator using a coupled equation
scheme, but drops the contributions from higher order Green's
functions.
This t-matrix is called $t_{pg}(Q)$, where
$Q=(i\Omega_{l},{\bf q})$ is a four-vector and $\Omega_{l}$ denotes the boson Matsubara frequency; it characterizes the non-condensed pairs which are
described physically and formally in different ways in the different
theories.
Here the subscript $pg$ is associated with the pseudogap (pg)
whose presence is dependent on the non-condensed or pre-formed
pairs.
Quite generally we can write the t-matrix in a ladder-diagram series
as $$ t_{pg}(Q) = \frac{U}{1 + U \chi(Q)},$$ where $\chi(Q)$ is the
pair susceptibility and $U$ denotes the attractive coupling constant.

The 
Nozi\`{e}res-Schmitt-Rink (NSR)
theory
\cite{NSR}
is associated with a pair susceptibility
$\chi(Q)$
which is a product
of two bare Green's functions.
The
fluctuation exchange or FLEX approach is associated
with two dressed Green's functions and
has been discussed by Haussmann \cite{Haussmann,
Zwerger}, Zwerger and their
collaborators in the context of the cold gases, and even
earlier in the context of the cuprates \cite{Tremblay2,Tchern,Micnas95}.
It is also called the
Luttinger-Ward formalism \cite{HaussmannRF}, or Galitskii-Feynman
theory \cite{Morawetz}.
Finally, it is well known
\cite{Kadanoff,Patton1971,Abrahams}
that BCS theory (and now
its BCS-BEC generalization) is associated with one bare and
one dressed Green's function in the pair susceptibility.

These differences would seem to be rather innocuous 
and technical but they have led to significant qualitative
differences and concurrently strong claims by various 
proponents.
We
stress that
while there are several variants, as we discuss below, the
version of the NSR scheme which seems to us most free of concerns is
that discussed in References~\cite{Strinaticuprates} which introduced
a more physical treatment of the number equation.  This revision of
strict NSR theory was, in part, an answer to J. Serene \cite{Serene}
who raised a question about a central approximation in the theory in
which the number equation ($n = 2 \sum_K G(K)$, where $G(K)$ is the
single particle Green's function) is approximated by
$n=-\frac{\partial\Omega_{th}}{\partial\mu}$, where the
thermodynamical potential $\Omega_{th}$ is approximated by a
ladder-diagram series. It was shown that this amounts to taking the
leading order in a Dyson series for $G(K)$.

The present paper concentrates on the normal state behavior,
although all three classes of theories have been extended
below $T_c$. What is essential about these extensions is that
the non-condensed pair excitations associated with
$t_{pg}$ are gapless, as in
boson theories. 
Indeed, it is in these $ T \leq T_c$
extensions that
a number of concerns have been raised.
In particular, in the leading order extended NSR theory (or so called
``Bogoliubov level" approach),
\cite{Strinati4,Drummond3,Randeriaab},
the gap
equation (which is assumed to take the usual BCS
form, rather than derived, for example, 
variationally) does not contain explicit pairing fluctuation
contributions; these enter indirectly only via the fermion chemical
potential $\mu$. At this level, the number equation is the only way in which
explicit pairing
fluctuations are incorporated.
At the so called ``Popov level" calculation, the 
gap equation is presumed to contain pair fluctuations
\cite{PS05}
but there is some complexity in ensuring the concomitant gaplessness of
the pair excitations. Similar issues arise with the FLEX
or Luttinger-Ward approach in which (\cite{HaussmannRF} and references therein) 
gapless sound modes
must
be imposed somewhat artificially.

While the order of the transition at $T_c$ is second order in
the BCS-Leggett scheme it is first order \cite{firstordertransitionpapers,Drummond3}
in NSR based approaches
(as well as for the fully renormalized pair susceptibility
scheme. This leads to unwanted features in the density profiles
\cite{Strinati4} and $T$ dependent superfluid density \cite{Griffingroup2},
$\rho_s(T)$.
Despite these unphysical aspects,
the NSR-based scheme captures the physics of Bogoliubov theory of weakly interacting bosons \cite{Strinati4}
and should, in principle,
be the quantitatively better low $T$ state, particularly in the
BEC limit. Nevertheless some issues have been identified \cite{Lerch}
which suggest the breakdown of true quasi-particles associated with
Bogoliubov-like theories for paired fermions. This, in turn,
derives from
the self consistent treatment of coupling between the non-condensed
pairs and the sound modes. Further analysis will be required to
establish if this is compatible with experimental or
theoretical constraints.

A very early concern about the so-called ``GG"
or FLEX approach was raised in a paper by Kadanoff and Martin \cite{Kadanoff}
in 1961:
``The similarity [to a Bethe-Salpeter equation] has
led several people to surmise that the symmetrical
equation [involving fully dressed G's everywhere] solved in
the same approximation would be more accurate. This surmise is
not correct. The Green's functions resulting from that equation
can be rejected in favor of those used by BCS by means of
a variational principle." 
Importantly this
approach does not have a true pseudogap. 
Despite claims by the Zwerger group \cite{HaussmannRF}
that theirs is a more fully ``consistent" theory,
and in this context appealing to Ref.~\cite{Moukouri},
the authors of Ref.~\cite{Moukouri} instead say:
``We thus conclude that ... approaches such as FLEX are
unreliable in the absence of a Migdal theorem and that there is
indeed a pseudogap."
Similar observations have appeared elsewhere in the literature
\cite{Moukouri,Tremblay2,Fujimoto,Micnas95}. 
As noted in Ref.~\cite{Fujimoto} `` vertex corrections to the self
energy, which are discarded in the previous studies [of FLEX] are
crucially important for the pseudogap".  Additional concerns have been
noted recently \cite{Morawetz} that in the FLEX (or $GG$ t-matrix) theory the propagator $G$ does not display
quasiparticle poles associated with
the gap. ``This is because the Dyson equation,
$G(k) = 1/(z-{\bf k}^2/2m-\Sigma(k))$, excludes identical poles
of $G$ and $\Sigma$ while the linear relation demands them''.

In recent work below $T_c$ 
\cite{Drummond2,Drummond3,Randeriaab}
a non-variational gap equation was used to derive an additional term in the number equation related to $\partial
\Omega_{th} / \partial \Delta_{sc} \neq 0$. Here $\Delta_{sc}$ is the
order parameter and, here, again, $\Omega_{th}$ is the thermodynamical
potential.  This extra term means there is no variational free energy
functional, such as required by Landau-Ginsburg theory. Of concern are
arguments that by including $\partial \Omega_{th} / \partial
\Delta_{sc} \neq 0$, it is possible to capture the results of Petrov et
al \cite{Petrov} for the inter-boson scattering length.  We see no
physical connection between the exact four-fermion calculations and
the non-variational component of the many body gap equation.  It
should, moreover, be stressed that all other t-matrix schemes have
reported an effective pair-pair scattering length given by $a_B=2a$
which is larger than the value $a_B=0.6a$ obtained from a four-body
problem \cite{Petrov}. Here $a$ is the $s$-wave scattering length of
fermions. Indeed, our past work \cite{Shina2} and that of Reference
\cite{PS05} have shown that one needs to go beyond the simple t-matrix
theory to accommodate these four-fermion processes. 

Additional concerns arise from the fact that an NSR-based scheme has
difficulty \cite{Parish,Hupolarized} accommodating polarization effects in the
unitary regime. As stated by the co-workers in Reference
\cite{Hupolarized}: ``Unfortunately, in a region around the unitary
limit we find that the NSR approach generally leads to a negative
population imbalance at a positive chemical potential difference
implying an unphysical compressibility.".  

The central weakness of the BCS-Leggett approach (and its finite-$T$
extension) appears to be the fact it focuses principally on the
pairing channel and is not readily able to incorporate Hartree
effects.  The evident simplicity of this ground state has raised
concern as well.  Clearly, this is by no means the only ground state
to consider but, among all alternatives, it has been the most widely
applied by the cold gas community including the following notable
papers
\cite{Stringari,Stringaricv,Randeria2,Cote,SR06,Rice2,Kinnunen,Machida2,BECBCSvortex,StrinatiJosephson,Basu}.
The central strengths of the finite-$T$ extended BCS-Leggett
approach in comparison with others
are that (i) there
are no spurious first order transitions and (ii) the entire range of
temperatures is accessible.
(iii) Moreover, polarization effects may be readily included
\cite{ChienPRL,heyan}, 
(iv) as may
inhomogeneities which are generally treated using
Bogoliubov deGennes theory \cite{BECBCSvortex,StrinatiJosephson},
based on this ground state.

The above analysis leaves us with two theoretical schemes which we wish
to further explore: the NSR approach (which in the normal phase
follows directly from the original paper \cite{NSR}) and
the BCS-Leggett-based scheme, as extended away from
zero temperature, and in particular above
$T_c$.
As t-matrix approaches to many body theory,
these are similar in spirit, but different
in implementation.
It is clearest below $T_c$, that the two theories focus on different physics.
NSR approaches view the dominant processes as the coupling
of the order parameter collective modes to the non-condensed
pairs and the BCS-Leggett scheme focuses on the steady
state equilibrium between the gapped fermions and the non-condensed
pairs.
Thus NSR focuses more fully on the bosonic degrees
of freedom and BCS-Leggett focuses on
the fermionic degrees of freedom.
Above $T_c$, because the NSR scheme involves only bare
Green's functions, it is simpler. Thus, it has been studied
at a numerical level in a more systematic fashion. In the
literature, the BCS-Leggett
approach at $T \neq 0$, 
has been addressed numerically \cite{Maly1,Maly2,Marsiglio,Morawetz}, 
assessed more theoretically \cite{Tremblay2}, 
as well as applied to different physical contexts 
\cite{Torma1,Torma2,Micnaslattice}. 
In this paper we apply an approximation based on prior
numerical work \cite{Maly1,Maly2} to simplify the calculations.

\subsubsection{The Fermionic Spectral Function}

A central way of characterizing these different BCS-BEC
crossover theories is through the behavior of the fermionic
spectral function, $A ({\bf k}, \omega)$. For the most part, here,
we restrict our consideration to the normal
state where $A ({\bf k}, \omega)$ should indicate the presence 
or not of a pseudogap. A momentum integrated form
of the spectral function is reflected in
radio frequency studies-- both tomographic \cite{MITtomo}
or effectively homogeneous and trap averaged
\cite{Grimm4,KetterleRF}.
One of the principal observations of this paper is that these
momentum integrated probes are not, in general, sufficiently sensitive to pick up
more than gross differences between the three crossover theories.

However, there are now momentum resolved RF studies \cite{Jin6} which
probe the spectral function more directly, in a fashion similar to
angle-resolved photoemission spectroscopy (ARPES) probes of condensed
matter.  A central aim of this paper is to show how these studies in
future will be able to differentiate more clearly between the
different crossover schools.  Here we confine our attention to
homogeneous systems, although experiments are necessarily done for the
trapped case. In addition to RF spectroscopy, it was proposed
\cite{GeorgesSpectral} that the spectral function can also be measured
in Raman spectroscopy.

We note that for the HTSCs, ARPES studies have been centrally important
in revealing 
information about the
superconducting as well as the pseudogap phases
\cite{arpesstanford_review}. 
Indeed, the close relation between ARPES and radio frequency probes 
has been discussed in our recent review \cite{RFlong}.
It was shown in Ref.~\cite{ANLPRL} that the spectral function of HTSCs
in the pseudogap phase appears to exhibit
dispersion features
similar to those 
in the superconducting phase. 
This spectral function is modeled \cite{Norman98,Maly1,Maly2}
by a broadened BCS form with
a self energy
\begin{equation}
\Sigma_{pg}(K) \approx
\frac{\Delta_{pg}({\bf k})^2}{\omega+\epsilon_k-\mu+i\gamma}
\label{eq:2a}
\end{equation}
Here $\Delta_{pg}({\bf k})$ is the ($s$ or $d$-wave) excitation gap of
the normal phase and $\gamma$ is a phenomenological damping.
Frequently, one adds an additional, structureless imaginary damping term
$i \Sigma_0$, as well.
High temperature superconductor experiments at temperatures 
as high as $T \approx 1.5 T_c$ have reported
that in the regions of the Brillouin zone (where the pseudogap is
well established), the dispersion of the fermionic excitations
behaves like
\begin{equation}
E_{\bf k} \approx \pm \sqrt{ (\epsilon_{\bf k} - \mu) ^2 + \Delta_{pg}({\bf k})^2}
\label{eq:3a}
\end{equation}

Importantly Eq.~(\ref{eq:3a}) has also been used in the cold gas
studies \cite{Jin6} in the region near and above $T_c$ and implemented
phenomenologically below $T_c$ \cite{GeorgesSpectral}. This both
demonstrates the presence of pairing and ultimately provides
information about the size of the pairing gap. It has been shown that
Eq.~(\ref{eq:2a}) is reasonably robust in the extended BCS-Leggett
state above $T_c$, at least up to temperatures \cite{Maly1,Maly2} of
the order of $\approx 1.3T_c$.  By contrast this approximate self
energy is not generally suitable to NSR theory
\cite{Maly1,Maly2}, although for $T/T_c = 1.001$ a fit to 
Eq.~(\ref{eq:3a})
has been obtained.  In a similar context we note that in the FLEX
approach, the spectral function and associated self energy is not of
the broadened BCS form. Mathematically, this BCS-like structure in the
self energy and fermionic dispersion (which is numerically obtained)
comes from the facts that the effective pair chemical potential
$\mu_{pair} \rightarrow 0$ at and below $T_c$, and that by having one
bare and one dressed Green's function in $\chi(Q)$ there is a gap in
the pair excitation spectrum so that the pairs are long lived; in this
way $\gamma$ (which scales with the inverse pair lifetime) is
small. Physically, we can say that this behavior reflects the
stability of low momentum pairs near $T_c$ and below.

These differences between the three different crossover
theories become less apparent for the 
momentum integrated
RF signals. In the BCS-Leggett approach at low temperatures
the dominant structure comes from pair
breaking of the condensate (which would be associated with the negative
root in Eq.~(\ref{eq:3a})). Despite the fact that their
fermionic dispersions are different, both other theories yield a very similar
``positive detuning branch" in the RF spectrum \cite{StoofRF,HaussmannRF}.
However, at higher temperatures, both for polarized and unpolarized gases,
there is theoretical
evidence of the ``negative detuning branch" arising from the
positive root in Eq.~(\ref{eq:3a}) in the BCS-Leggett
based approach \cite{momentumRF,RFlong}. This is absent in the two other
schemes, at least within the normal state. It also appears to be difficult
to see experimentally in the unpolarized case, although it is clearly
evident once even a small polarization \cite{Rice2,KetterleRF}
is present \cite{MITtomoimb}.

This paper is organized as follows. Sections~\ref{sec:NSRtheory} and
\ref{sec:G0Gtheory} briefly review NSR theory and BCS-Leggett theory
as extended to non-zero $T$.  Section~\ref{sec:Spectral} addresses a
comparison of the spectral function at unitarity and on the BEC side
obtained from the two theories. In subsections ~\ref{sec:DOS} we plot
a comparison of the related density of states at unitarity and in
~\ref{sec:RF} we address a comparison of RF spectra in the two
theories for an unpolarized Fermi gas which also addresses
experimental data. Also included is a prediction of RF spectra on the
BEC side of resonance. The remaining sections (Section~\ref{sec:RFTc}
and Section~\ref{sec:RFpol}) do not focus on comparisons because the
issues discussed pertain to questions which only BCS-Leggett theory
has been able to address. Here we propose a subtle signature of the
superfluid transition in Section~\ref{sec:RFTc} which could be
addressed in future and in Section~\ref{sec:RFpol} we address the
theoretical RF spectrum of polarized Fermi gases at unitarity and its
comparison with experimental data. Section~\ref{sec:conclusion}
concludes this paper.

We remark that in this paper we study $s$-wave pairing in three
spatial dimensions which is more relevant to ultra-cold Fermi gases
while HTSCs should be modeled as $d$-wave pairing in quasi-two
dimensions. However, as one will see, there are many interesting
common features in these two systems.

\section{NSR Theory Above $T_c$}
\label{sec:NSRtheory}

The normal state treatment of NSR theory which we apply here
follows directly from the original paper
in  Ref.~\cite{NSR}. Here the different variants of NSR theory
as introduced by different groups \cite{Drummond3,Drummond5,Strinati4,PS05}
are not as important.
Although there is still the concern \cite{Serene}
that the number equation is only approximate, the numerics
are simpler if we follow the original approach; comparisons
with more
recent work in Ref.~\cite{OhashiNSR} 
(based on the fully consistent number equation
($n=2\sum_{K}G(K)$)
seem to validate this
simplification.
This same more consistent number equation is used throughout the
work by the Camerino group \cite{Strinaticuprates}.

NSR theory builds on the fact that the fermion-fermion attraction
introduces a correction to the thermodynamic potential:
$\Delta\Omega_{th}=\Omega_{th}-\Omega_{f}=\sum_{Q}\ln[U^{-1}+\chi_{0}(Q)]$,
where $\Omega_{f}$ is the thermodynamic potential of a non-interacting
Fermi gas, $\chi_{0}(Q)$ is the NSR pair susceptibility, and
$\sum_{Q}=T\sum_{l}\sum_{\bf q}$. In the normal phase,
\begin{eqnarray}
\chi_{0}(Q)&=&\sum_{K}G_{0}(K)G_{0}(Q-K) \nonumber \\
 &=&\sum_{\mathbf{k}}\frac{f(\xi_{\mathbf{k}+\mathbf{q}/2})+f(\xi_{\mathbf{k}-\mathbf{q}/2})-1}{i\Omega_{l}-(\xi_{\mathbf{k}+\mathbf{q}/2}+\xi_{\mathbf{k}-\mathbf{q}/2})}.
\end{eqnarray}
Here $K=(i\omega_{\nu},{\bf k})$ where $\omega_{\nu}$ is the fermion
Matsubara frequency, $\sum_{K}=T\sum_{\nu}\sum_{\bf k}$,
$G_{0}(K)=1/(i\omega_{\nu}-\xi_{\bf k})$ is the non-interacting
fermion Green's function, $\xi_{\bf k}=k^2/2m-\mu$ where $m$ and $\mu$
denote the mass and chemical potential of the fermion, and $f(x)$ is
the Fermi distribution function. We set $\hbar\equiv 1$ and
$k_{B}\equiv 1$. 
NSR theory is constrained by the condition
$U^{-1}+\chi_{0}(0)>0$ since $U^{-1}+\chi_{0}(0)=0$ signals an
instability of the normal phase and the system becomes a superfluid as
temperature decreases.

The fermion chemical potential is determined
via the NSR number equation 
\begin{equation}\label{eq:NSRneq}
n=-\partial\Omega_{th}/\partial\mu,
\end{equation} 
where $n$ is the total fermion
density. 
As noted in \cite{NSR}, when $\mu<0$,
$U^{-1}+\chi_{0}(i\Omega_{l}\rightarrow \Omega+i0^{+},{\bf q})=0$ has
solutions which correspond to bound states. Those bound states have
contributions proportional to $b(\Omega^{(0)}_{\bf q})$ to the total
density, where $b(x)$ is the Bose distribution function and
$\Omega^{(0)}_{\bf q}$ is the energy dispersion of the bound
states.

\begin{figure} 
\includegraphics[width=3.in,clip] 
{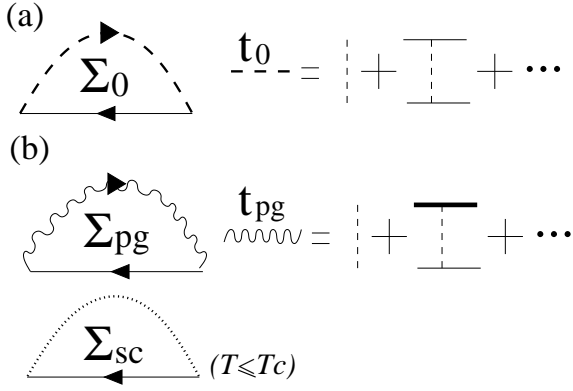}
\caption{Fermion self-energy and ladder diagrams in (a) NSR theory and (b) $GG_0$ t-matrix theory. The thin solid line, thick solid line, thick dashed line, dotted line, wavy line, vertical thin dashed line denote  $G_0$, $G$, $t_0$, $t_{sc}=-\Delta_{sc}^{2}\delta(Q)/T$, $t_{pg}$, $U$.}
\label{fig:SE} 
\end{figure} 

The NSR Green's function is $G(K)=[G_{0}(K)-\Sigma_{0}(K)]^{-1}$ 
and its retarded form is $G_{R}(\omega,{\bf k})
=G(i\omega_{n}\rightarrow\omega+i0^{+},{\bf k})$.
Following a t-matrix formalism, one can also consider the corrections
to the fermion self-energy
$\Sigma_{0}(K)=\sum_{Q}t_{0}(Q)G_{0}(Q-K)$. Figure~\ref{fig:SE}
illustrates the structure of fermion self-energy in NSR theory and in
the finite temperature theory associated with 
the BCS-Leggett ground state, which will be summarized in the next
section. Here the t-matrix is given by
$t_{0}(Q)=1/[U^{-1}+\chi_{0}(Q)]$. The retarded form of the fermion
self-energy has the structure $\Sigma_{0}(i\omega_{\nu}\rightarrow
\omega+i0^{+},{\bf k})=\Sigma_{0}^{\prime}(\omega,{\bf
k})+i\Sigma_{0}^{\prime\prime}(\omega,{\bf k})$, where
$\Sigma_{0}^{\prime}$ and $\Sigma_{0}^{\prime\prime}$ correspond to
the real and imaginary part of the self-energy. We separate the
contribution of the bound states from the rest (called the continuum
contribution). The continuum contribution is
\begin{eqnarray}
\Sigma_{0c}^{\prime}(\mathbf{k},\omega)&=&\sum_{\mathbf{q}}\Big\{f(\xi_{\mathbf{q}})\mbox{Re}[t_{0R}(\mathbf{q}+\mathbf{k},\omega+\xi_{\mathbf{q}})]+ \nonumber \\
& &\mathcal{P}\int_{-\infty}^{\infty}\frac{d\Omega}{\pi}\frac{b(\Omega)}{\Omega-\omega-\xi_{\mathbf{q}}}\mbox{Im}[t_{0R}(\mathbf{q}+\mathbf{k},\Omega)]\Big\}, \nonumber \\
\Sigma_{0c}^{\prime\prime}(\mathbf{k},\omega)&=&\sum_{\mathbf{q}}[b(\omega+\xi_{\mathbf{q}})+f(\xi_{\mathbf{q}})]\mbox{Im}[t_{0R}(\mathbf{q}+\mathbf{k},\omega+\xi_{\mathbf{q}})]. \nonumber \\
\end{eqnarray}
Here $t_{0R}(\Omega,{\bf q})=t_{0}(i\Omega_{l}\rightarrow\Omega+i0^{+},{\bf q})$ and $\mathcal{P}$ denotes the Cauchy principal 
integral. In the presence of bound states, $t_{0R}(\Omega,{\bf q})$ has poles which result in bound state contributions to the fermion self-energy 
\begin{eqnarray}
\Sigma_{0b}^{\prime}(\mathbf{k},\omega)&=&-\mathcal{P}\sum_{\mathbf{q}}b(\Omega_{bs})\frac{1}{\frac{\partial\mbox{Re}[t_{0R}^{-1}]}{\partial\Omega}\Big|_{\Omega_{bs}}}\left[\frac{1}{\Omega_{bs}-\omega-\xi_{\mathbf{q}-\mathbf{k}}}\right] \nonumber \\
\Sigma_{0b}^{\prime\prime}(\mathbf{k},\omega)&=&-\sum_{\mathbf{q}}\pi b(\Omega_{bs})\frac{1}{\frac{\partial\mbox{Re}[t_{0R}^{-1}]}{\partial\Omega}\Big|_{\Omega_{bs}}}\delta(\Omega_{bs}-\omega-\xi_{\mathbf{q}-\mathbf{k}}). \nonumber \\
\end{eqnarray}
Here $\Omega_{bs}=\Omega_{bs}({\bf q})$ denotes the location of the pole in $t_{0R}$.


\section{BCS-Leggett Theory: Broken Symmetry Phase}
\label{sec:G0Gtheory}

We first review BCS-Leggett theory as it has been applied in the
broken symmetry phase.
The first three equations below represent a t-matrix
approach to the derivation of the \textit{standard} BCS gap equation.
In this way we set up a machinery which is readily
generalized to include BCS-BEC crossover theory.
BCS theory can be viewed as
incorporating \textit{virtual} non-condensed pairs.
Because they
are in
equilibrium with the condensate, the non-condensed pairs
must have a vanishing
``pair chemical potential", $\mu_{pair} =0$. Stated alternatively
they must be gapless. The t-matrix can be derived from the ladder diagrams in the particle-particle channel (see Fig.~\ref{fig:SE}):
%
\begin{equation}
t_{pg} (Q) \equiv \frac {U} { 1 + U \sum_{K} G(K) G_0(-K+Q)},
\label{eq:3}
\end{equation}
with $t_{pg}(Q=0)\rightarrow \infty$, which is equivalent to $\mu_{pair} =0$, for $T \leq T_c$. Here $G$, and $G_0$ represent dressed and
bare Green's functions, respectively.
To be consistent with the BCS ground state of Eq.~(\ref{eq:1}), 
the self energy is
\begin{eqnarray}
\Sigma_{sc} (K)&=& \sum_Q t_{sc}(Q) G_0(-K+Q) \nonumber \\
&=&-\sum_Q \frac {\Delta_{sc}^2}{T} \delta(Q) G_0 (-K+Q) \nonumber \\
&=& -\Delta_{sc}^2 G_0(-K).
\label{eq:4}
\end{eqnarray}
$\Delta_{sc}(T)$ is the
order parameter while $\Delta(T)$ is the pairing gap.
From this one can write down the full Green's function, $G(K)=[G_{0}^{-1}(K)-\Sigma_{sc} (K)]^{-1}$. Finally, Eq.~(\ref{eq:3}) 
with $\mu_{pair} = 0$ gives the BCS gap equation below $T_c$: 
\begin{equation}
1 = -U \sum_{\bf k}
\frac{1 - 2 f (E_k^{sc})} { 2 E_k^{sc}}
\label{eq:5}
\end{equation}
with 
$E_k^{sc} \equiv \sqrt{ (\epsilon_k- \mu)^2 + \Delta_{sc}^2}$. 
We have, thus, used Eq.~(\ref{eq:3}) to
derive the standard BCS gap equation within a t-matrix
language and the result appears in Eq.~(\ref{eq:5}).
Eq.~(\ref{eq:3}) above can be viewed as representing
an extended version of the Thouless criterion of strict BCS which applies
for all $ T \leq T_c$.
This derivation
leads us to reaffirm the well known result \cite{Kadanoff,Patton1971,Abrahams}
that BCS theory 
is associated with one bare and
one dressed Green's function in the pair susceptibility.

Next, \textit{to address BCS-BEC crossover, we
feed back the contribution of the non-condensed pairs
which are no longer virtual as they are in strict BCS theory}, above.
Eq.~(\ref{eq:3}) is taken as a starting point. Equation~(\ref{eq:4}) is revised
to accommodate this feedback.
Throughout,
$K,Q$ denote four-vectors.
\begin{widetext}
\begin{equation}
\Sigma(K) = \sum_{Q} t(Q) G_0 (-K + Q) = \sum_Q [t_{sc}(Q) + t_{pg}(Q) ]
G_0 (-K + Q)  = \Sigma_{sc}(K) + \Sigma_{pg}(K)
\label{eq:6}
\end{equation}
\begin{equation}
\mbox{Numerically},~
\Sigma_{pg}(K) \approx
\frac{\Delta_{pg}^2}{i\omega_{\nu}+\epsilon_{\bf k}-\mu+i\gamma}+i\Sigma_{0};
~\mbox{analytically}, ~~\Sigma_{sc}(K) =
\frac{\Delta_{sc}^2}{i\omega_{\nu}+\epsilon_{\bf k}-\mu}.
\label{eq:9}
\end{equation}
\begin{equation}
\gamma,\Sigma_{0} ~ \mbox{ small: } \Sigma(K) \approx - (\Delta_{sc}^2 + \Delta_{pg}^2) G_0(-K) \equiv
- \Delta^2 G_0(-K) ~\Rightarrow
\Delta_{pg}^2 \equiv -\sum_Q t_{pg}(Q)
\label{eq:7}
\end{equation}
\begin{equation}
t_{pg} (Q=0) = \infty \Rightarrow~~ 1 = -U
\sum_{\bf k} \frac{1 - 2 f(E_{\bf k})}{2 E_{\bf k}},~~~
E_{\bf k} \equiv \sqrt{ (\epsilon_{\bf k}- \mu)^2 + \Delta^2},~~
\label{eq:8}
\end{equation}
\end{widetext}
Note that
Eqs.~(\ref{eq:4}) and (\ref{eq:6}) introduce the self energy which is
incorporated into the fully dressed Green's function $G(K)$, appearing
in $t_{pg}$. Also note the number equation $n = 2 \sum_K G(K)$ is to
be solved consistently:
\begin{equation}
n = 2 \sum_K G(K) = \sum _{\bf k} \left[ 1 -\frac{\xi_{\bf k}}{E_{\bf k}}
+2\frac{\xi_{\bf k}}{E_{\bf k}}f(E_{\bf k})  \right]
\label{eq:neq}
\end{equation}
where $\xi_{\bf k} = \epsilon_{\bf k} - \mu$.

This leads to a closed set of equations for the pairing gap $\Delta(T)$,
and the pseudogap $\Delta_{pg}(T)$ (which can be derived from Eq.~(\ref{eq:7})). The BCS-Leggett approach with the dispersion shown in Eq.~(\ref{eq:8}) thus provides a microscopic derivation for the pseudogap model implemented in Ref.~\cite{GeorgesSpectral}. 
To evaluate $\Delta_{pg}(T)$ numerically, we assume the main contribution to $t_{pg}(Q)$ is from non-condensed pairs with small $Q$, which is reasonable if temperature is not too high \cite{Maly1,Maly2}. The inverse of $t_{pg}$ after analytical
continuation is approximated as $t_{pg}(\omega,{\bf q})\approx [Z(\Omega -
\Omega^0_{\bf q}+\mu_{pair}) + i \Gamma^{}_Q]^{-1}$, where
$Z=(\partial\chi/\partial\Omega)|_{\Omega=0,q=0}$, $\Omega^0_{\bf
q}=q^2/(2M_{b})$ with the effective pair mass
$M_{b}^{-1}=(1/3Z)(\partial^{2}\chi/\partial q^{2})|_{\Omega=0,q=0}$
which takes account of the effect of pair-pair interactions. Near
$T_c$, $\Gamma^{}_Q \rightarrow 0$ faster than $q^2$ as $q\rightarrow
0$.  Following this approximation, $\Delta_{pg}(T)$ essentially
vanishes in the ground state where $\Delta = \Delta_{sc}$.

The entire derivation contains one simplifying
(but not fundamental) approximation. Detailed numerical calculations
\cite{Maly1,Maly2} show that
$\Sigma_{pg}$ can be written as in Eqs.~(\ref{eq:2a}), which is the
same as that in Eq.~(\ref{eq:9}), with the
observation that as
$T_c$ is approached
from above, $\gamma$ and $\Sigma_{0}$ which appears in Eq.~(\ref{eq:9})
become small.
To zeroth
order, then,
we drop
$\gamma$ and $\Sigma_{0}$ 
(as in Eq.~(\ref{eq:7})), and thereby can more readily solve the gap equations.
To first order we include this lifetime effect
as in Eq.~(\ref{eq:9}) in addressing  spectral functions and
other correlations.

The actual value of $\gamma$ makes very little
qualitative difference and the previous numerical calculations
\cite{Maly1,Maly2} do not include
$d$-wave or trap effects so that we should
view $\gamma$ as a phenomenological parameter.
For the HTSCs,
the expression for $\Sigma_{pg}$ in Eq.(\ref{eq:2a}) is
standard in the field \cite{Norman98,Normanarcs,FermiArcs},
and we can use specific
heat jumps or angle resolved photoemission to deduce $\gamma$,
as others \cite{Norman98} have done.
For the cold gases
the
precise value of $\gamma$, and its $T$-dependence are not particularly
important, as long as it is non-zero at
finite $T$. 
In this paper we will deduce reasonable values for
$\gamma(T)$ and $\Sigma_{0}$ from tomographic RF experiments.

\subsection{Extension Above $T_c$}

We can expand $t_{pg}(Q)$ at small $Q$, and in the normal state
to find
\begin{equation}
t_{pg}^{-1} (0) \equiv Z \mu_{pair} = U^{-1} + \chi(0)
\end{equation}
where the residue $Z$ and pair dispersion (not shown) $\Omega_q$,
are then determined \cite{heyan2}.
This is associated with the normal state gap equation
\begin{equation}
U^{-1} + \sum_{\bf k}
\frac{1-2 f(\Ek)}{2 \Ek}= Z\mu_{pair}  \,,
\label{eq:pggap}
\end{equation}
Similarly, above $T_c$, the pseudogap contribution to $\Delta^2(T) =
{\Delta}_{sc}^2(T) + \Delta_{pg}^2(T)$ is given by
\begin{equation}
\Delta_{pg}^2=\frac{1}{Z} \sum_{\bf q}\, b(\Omega_q -\mu_{pair}) \,\,.
\label{eq:1a}
\end{equation}
The number equation remains unchanged.
In summary, when the temperature is above $T_c$, the order parameter is
zero, and $\Delta=\Delta_{pg}$. Since there is no condensate,
$\mu_{pair}$ is nonzero, and the gap equation is modified.
From these
equations, one can determine $\mu$,
$\Delta$ and $\mu_{pair}$.

\subsection{Incorrect Criticism from the Drummond Group}

The Drummond group \cite{Drummond5} has made a number of
incorrect claims about our past work which we address here.
The authors claim to have numerically studied the behavior
associated with the three possible
pair susceptibilities.
We note that there is no elemental numerical data in their paper,
nor do they present details beyond their use of an
``adaptive step Fourier transform"
algorithm.
This should be compared with
work by the Tremblay group
\cite{Moukouri,Tremblay2,Tremblay3}
and others \cite{Fujimoto,Marsiglio}.  It is hoped that in future they
will present plots of the t-matrix and self energy to the community,
to the same degree that we have shared the output of our numerical
schemes in References \cite{Maly1} and \cite{Maly2}.  Important will
be their counterparts to Figs. 8a and 9 (lower inset) in
Ref.~\cite{Maly1}, which show how reliable the form in
Eq.~(\ref{eq:9}) is for the full $GG_0$ self energy.  More
specifically: they have argued that the ``decomposition into $pg$ and
$sc$ contributions [see Eq.~(\ref{eq:6}) above], omits important
features of the full theory". This claim is incorrect and is based on
their Fig.~1 of Ref.~\cite{Drummond5} which can be seen to be
unrelated to the pg and sc decomposition, since their analysis of our
so-called ``pseudogap theory" is confined to the normal
phase. \textit{The decomposition only applies below $T_c$}.
Moreover, we refute the argument that the decomposition
into sc and pg terms shown in Eq.~(\ref{eq:6}) above is
unphysical. This decomposition is associated with the
fact that there are necessarily both condensed and non-condensed
pairs in the Fermi gases at unitarity. This break-up
is standard in studies of Bose gases. The
details of how to describe the $pg$ contribution,
but not its necessary presence in a decomposed
fashion,
are what varies from theory to theory.

Importantly, the
``discrepancies" associated with thermodynamical
plots based on our approach should be attributed to
the absence of a Hartree term, not to any deeper physics.
The reader can see that if the usual $\beta$ parameter
is changed from $-0.41$
to around $-0.6$ the BCS-Leggett curve will be aligned with
the others.

\section{Comparisons of the Spectral Function}
\label{sec:Spectral}
\begin{figure} 
\begin{center}
\includegraphics[width=3.4in,clip] 
{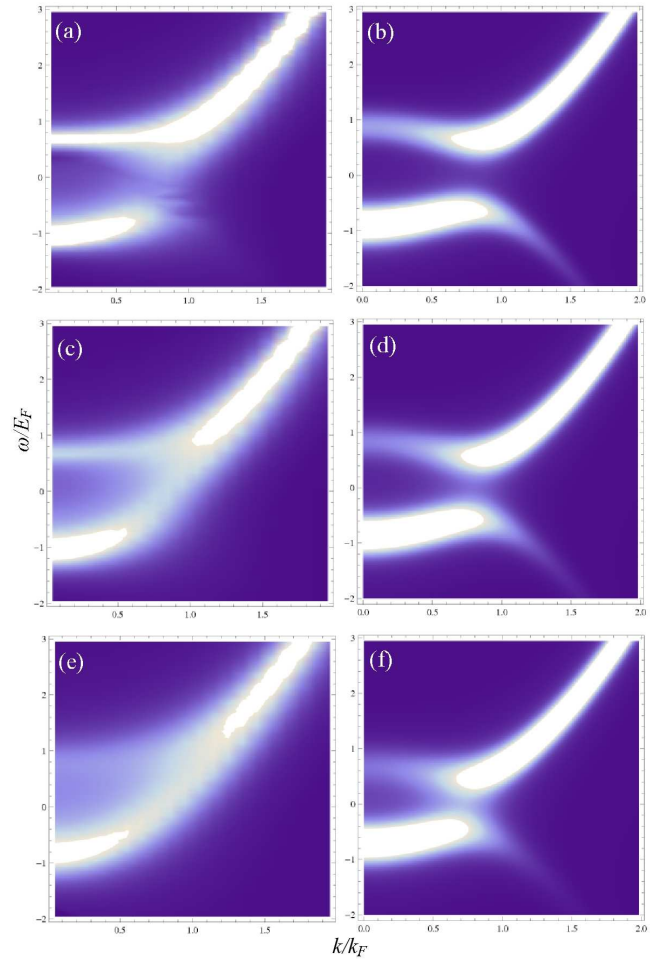}
\caption{(Color online) Spectral function obtained from NSR theory (left column) and from the extended BCS-Leggett theory (right column) at unitarity. Temperatures ($T/T_F$) from top to bottom are: (a) and (b) $0.24$, (c) and (d) $0.34$, (e) and (f) $0.55$. The ranges of $k/k_{F}$ and $\omega/E_F$ are $(0,2)$ and $(-2,3)$, respectively. }
\label{fig:Sp_Ia0} 
\end{center}
\end{figure}

The fermionic spectral function is given by $A(\omega,{\bf k})
=-2\mbox{Im}[G_{R}(\omega,{\bf k})]$, where $G_{R}$ is the retarded
Green's function. In this section we want to explore its behavior both
as a function of wavevector ${\bf k}$ and of frequency $\omega$. Of
particular interest is the question of whether there is a pseudogap in
the spectral function.  There are different criteria for arriving at
an answer. Importantly, depending of this choice the answer will be
different for NSR theory (and also, it appears for the FLEX theory of
Ref.~\cite{HaussmannRF}). The following definitions come from
different measurements of HTSCs which are not internally
contradictory.  Following Ref.~\cite{Timusk} and references therein,
we examine two criteria for the pseudogap in HTSCs.
\begin{enumerate}
\item One can define the existence of a pseudogap as 
associated with the observation that
$A ({\bf k}, \omega)$ as a function of $\omega$ at $ k = k_F$
exhibits a two-peak structure in the normal state. This definition is particularly useful for spectroscopies such as ARPES which can probe the spectral function near the Fermi surface.
\item Alternatively, the existence of a pseudogap can be identified when the density of states (DOS) (which represents
an integral over ${\bf k}$ of the spectral function) 
is depleted near the Fermi energy. This definition appeals to tunneling experiments where the DOS can be measured.
\end{enumerate}

In addressing these criteria it
is useful to refer to the spectral function of the BCS-Leggett 
ground state given by 
$A_{BCS}(\omega,{\bf
k})=u_{k}^{2}\delta(\omega-E_k)+v_{k}^{2}\delta(\omega+E_k)$, where
$u_{k}^{2},v_{k}^{2}=(1/2)[1\pm(\xi_{k}/E_{k})]$. As a function of
frequency, there are two
branches: the upper branch located at $\omega=E_k$ has weight
$u^{2}_{k}$ and the lower branch located at $\omega=-E_k$ has weight
$v^{2}_{k}$. Since $E_k\ge \Delta$, the spectral function is gapped at
all ${\bf k}$. One recognizes two features in $A_{BCS}$. First, there
is particle-hole mixing which results in the two branches. Second, there
is an upwardly dispersing and a downwardly dispersing symmetric
contribution to the spectral function arising from the $\pm$
signs in Eq.~(\ref{eq:3a}).
This is symmetric about 
the non-interacting Fermi energy. 
At finite temperatures, as one
will see, both NSR theory and the extended-BCS-Leggett theory show
particle-hole mixing in the sense that there are two
branches in the spectra. In contrast, the fermionic dispersion
in NSR theory does not lead to two symmetric upwardly and downwardly
dispersing branches.
The behavior of the finite $T$ spectral function associated with BCS-Leggett
theory, given in Eq.~(\ref{eq:2a}) is, however,
rather similar to its superfluid
analogue.

It is unlikely that Eq.~(\ref{eq:2a}) will be appropriate at sufficiently
high temperatures.
Indeed, one can see from Figure 3 in Reference \cite{Maly2} and
the surrounding discussion that numerical calculations show this
approximation is appropriate up to some temperatures of the order
of $T/T_c \approx 1.3$ for a system near unitarity and in the absence
of a trap.
In a fully consistent numerical calculation one expects that as
$T$ is raised the pseudogap will decrease so that the pair susceptibility
of the extended BCS-Leggett
theory should eventually evolve from $GG_0$ to $G_0G_0$. In this way
the fully numerical NSR scheme should be very reasonable at sufficiently
high $T$, where the pseudogap begins to break down. Physically
we can argue that the BCS-Leggett scheme is better suited to treating
pairs which have pre-dominantly low momentum, and thus, it should apply
closer to condensation.
For the purposes of comparison, in this section we apply Eq.~(\ref{eq:2a}) up
to somewhat higher temperatures than appears strictly feasible.

\begin{figure} 
\includegraphics[width=3.4in,clip] 
{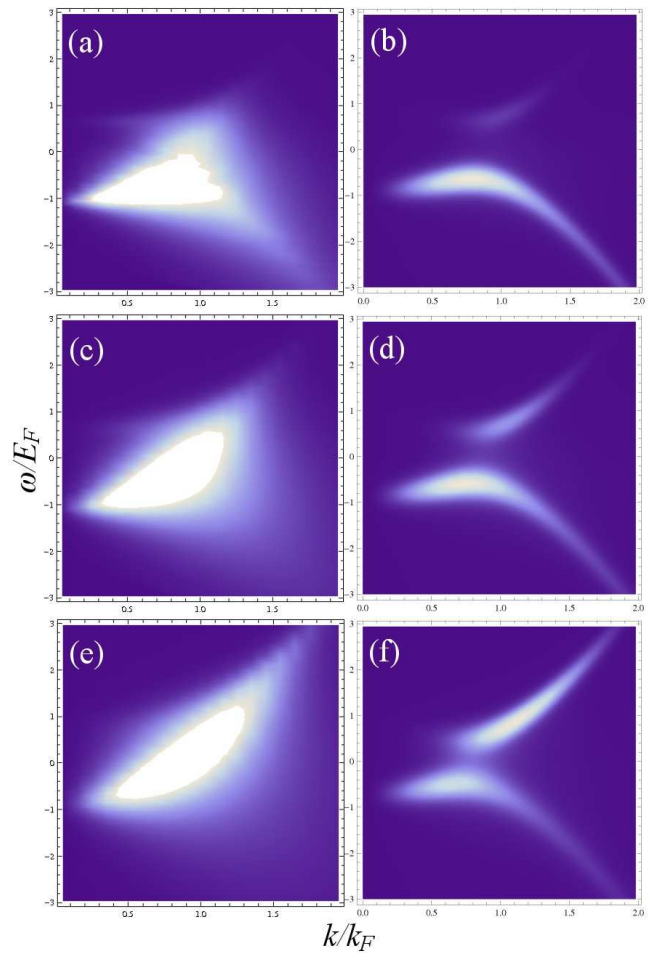}
\caption{(Color online) The plot of the function $k^{2}f(\omega)A({\bf k},\omega)/2\pi^{2}$ calculated from NSR theory (left column) and from the extended BCS-Leggett theory (right column) at $1/k_{F}a=0$. Temperatures ($T/T_F$) from top to bottom are: (a) and (b) $0.24$, (c) and (d) $0.34$, (e) and (f) $0.55$. The ranges of $k/k_{F}$ and $\omega/E_F$ are $(0,2)$ and $(-3,3)$, respectively. }
\label{fig:kRF_Ia0} 
\end{figure}

A very important physical distinction emerges between the different
models for the pair susceptibility which is then reflected in
the fermionic self energy and ultimately in the spectral function. 
Because a dressed Green's function appears in BCS-Leggett theory,
the t-matrix $t(Q)$ at small $q$
has a notably different behavior, particularly
at low $\omega$ as compared with the NSR case. This is seen
most clearly by comparing Figure 2 and Figure 9 in Ref.\cite{Maly1}.
This difference can be seen as a gap in the $GG_0$ t-matrix which
serves to stabilize the pair excitations. In the normal state
the pairs live longer when a pseudogap is present because of
this feedback. As a result the behavior of the fermionic self energy is
different, leading to a reasonable fit to Eq.~(\ref{eq:2a})
in $GG_0$ theory as shown by the lower inset in Figure 9
of Reference \cite{Maly1}, as compared with the poorer
fit to Eq.~(\ref{eq:2a}) found in NSR theory and shown in
Figure 8a from Ref. \cite{Maly1}.
We will reach qualitatively similar conclusions in the next
section of the paper.
We summarize by noting that the extended 
BCS-Leggett theory focuses on low $q$ pairs which
dominate near condensation.
NSR theory treats pairing without
singling out low $q$ only.
Each of these theories should be appropriate in different temperature
regimes of the normal state.
Concomitantly,
because of the enhanced stability of the pairs, the broadening of
the spectral peaks will be considerably smaller in BCS-Leggett theory
as compared with NSR theory.

To make a connection with experiments
on ultra-cold Fermi gases, we regularize the attractive coupling
constant via $U^{-1}=m/(4\pi a)-\sum_{\bf k}(m/k^{2})$.
We choose as our units the Fermi energy $E_F$, or, as appropriate
the Fermi temperature $T_F$, or
Fermi momentum $k_F$ of a non-interacting Fermi gas with the
same particle density.
The unitary point where $a$ diverges is of particular interest because
two-body bound states emerge. Since many-body effects renormalize the
coupling constant, the fermion chemical potential remains positive at
unitarity in both NSR theory and the BCS-Leggett theory. This implies
that bound states in a many-body sense have not fully
emerged. 
In our numerics,
we choose $\gamma(T)$ to be very roughly consistent
with RF experiments. For the unpolarized case we set
$\gamma/E_F=0.12(T/T_c)$ at unitarity and included a small background
imaginary term
$\Sigma_0/E_F=0.05$.

\subsection{Comparison of Spectral functions via Contour plots}

Figure~\ref{fig:Sp_Ia0} present a plot of the spectral function at
unitarity ($1/k_{F}a=0$) obtained from NSR theory (left column) and
from the BCS-Leggett t-matrix theory (right column) at selected
temperatures. 
In the BCS-Leggett case we use
the approximation that the self energy associated
with non-condensed pairs is of a broadened BCS form (in Eq.~(\ref{eq:2a})).
The transition temperatures
$T_c/T_F=0.238$ and $T_c/T_F=0.26$ are obtained for the
 NSR and BCS-Leggett cases respectively. 
Both theories yield higher $T_c$ values
than found \cite{QMCTc} in quantum Monte Carlo simulations, where
$T_c/T_F\approx 0.15$ at unitarity. The $T_c$ curves in BCS-BEC crossover of NSR and the extended BCS-Leggett theories are shown in Ref.~\cite{CompareReview}.

In Fig.~\ref{fig:Sp_Ia0} the comparisons are made at three different
temperatures. The horizontal and vertical axes on each panel
correspond to wave number and frequency and what is plotted in the
contour plots is the fermionic spectral function for a
three-dimensional homogeneous gas. The white areas correspond to peaks
in the spectral function and they map out a dispersion for the
fermionic excitations. With the possible exception of the highest $T$
NSR case (lower left figure) the spectral functions in all cases shown
in Fig.~\ref{fig:Sp_Ia0} are gapped at small $k/k_F$, which indicates
the existence of particle-hole mixing The lower branch of the spectral
function from the BCS-Leggett t-matrix theory clearly shows a downward
bending for $k>k_{F}$ which is associated with a broadened BCS-like
behavior. The spectral function of a phenomenological pseudogap model
presented in Ref.~\cite{GeorgesSpectral}, which \textit{can be derived
microscopically from the BCS-Leggett approach}, exhibits
similar contour plots as ours from the extended BCS-Leggett theory.
By contrast, in NSR theory the lower branch corresponds to very
broad and very small peaks when $k>k_{F}$ which are barely
observable. We see no clear evidence of a downward dispersing branch
even at the lowest temperature above $T_c$.

As a function of temperature, in the NSR case, the physics suggests a
smooth evolution with increasing $T$ towards a single branch, upwards
dispersing-- almost Fermi liquid dispersion.  It seems likely that as
$T$ is raised and the pairing gap becomes less important the pair
susceptibility of the BCS-Leggett state should cross from $GG_0$ to
$G_0G_0$ so that the two schemes merge. This means that our previous
simplification of the fermionic self energy $\Sigma$ (as a broadened
BCS form, see Eq.~(\ref{eq:9})) is no longer suitable in this high $T$
regime.  There are not really two different types of pseudogap, but
rather the extended BCS-Leggett theory represents pairs which are
close to condensation-- and thus predominantly low momentum
pairs. [This is built into the approximation Eq.~(\ref{eq:9}), which
was used to model the pseudogap self energy.]  By contrast the NSR
case considers pairs with a broad range of momenta.

One can see that the vanishing of the pseudogap as temperature
increases is different in the two theories. 
In the extended BCS-Leggett theory, the two
branches approach each other and the gap closes while in NSR theory,
the spectral function fills in the gapped regime and its overall shape
evolves toward a single parabola. 
We note that 
our spectral function from NSR theory at unitarity looks identical to the results in Ref.~\cite{OhashiNSR}, even though the latter was computed with a more
self consistent number equation.
This
implies that the difference between the two number equations (Eq.~(\ref{eq:NSRneq}) and $n=2\sum_{K}G(K)$) has no qualitative impact. 
Interestingly, the spectral function at the highest $T$ 
from NSR theory resembles that presented in Ref.~\cite{HaussmannRF} for the
near-$T_c$ normal phase of $GG$ t-matrix theory.

It should be noted that
this spectral function is not the quantity directly measured in
momentum resolved RF studies \cite{Jin6}. Rather what is measured
there is the function 
$k^{2}f(\omega)A({\bf k},\omega)/2\pi^{2}$ which is plotted in
Figure~\ref{fig:kRF_Ia0}. This convolution preserves the large momentum
part of the lower branch of the spectral function and suppresses the
remainder. The downward bending behavior is clearly observed in the extended BCS-Leggett theory again (right column) for all three temperatures. In contrast, 
only the lowest
temperature plot of NSR theory ($T/T_F=0.24$, slightly above $T_c$)
shows a weak downward bending at large momentum. This downward
dispersion,
however, cannot be observed at higher temperatures (left column). In 
the actual momentum-resolved RF experiments of trapped Fermi
gases \cite{Jin6} trap averages enter so that the actual curves
are substantially broader \cite{momentumRF}.
A clear signature of downward dispersion in
experimental data will help determine whether the pseudogap phase with
noncondensed pairs behaves in a way which is similar to
the HTSCs, where this feature has been reported \cite{ANLPRL}.
It should be noted that it is this signature which has been
used in Ref.~\cite{Jin6} to arrive at an indication of
the presence of pairing.

\begin{figure}
\includegraphics[width=3.4in,clip]
{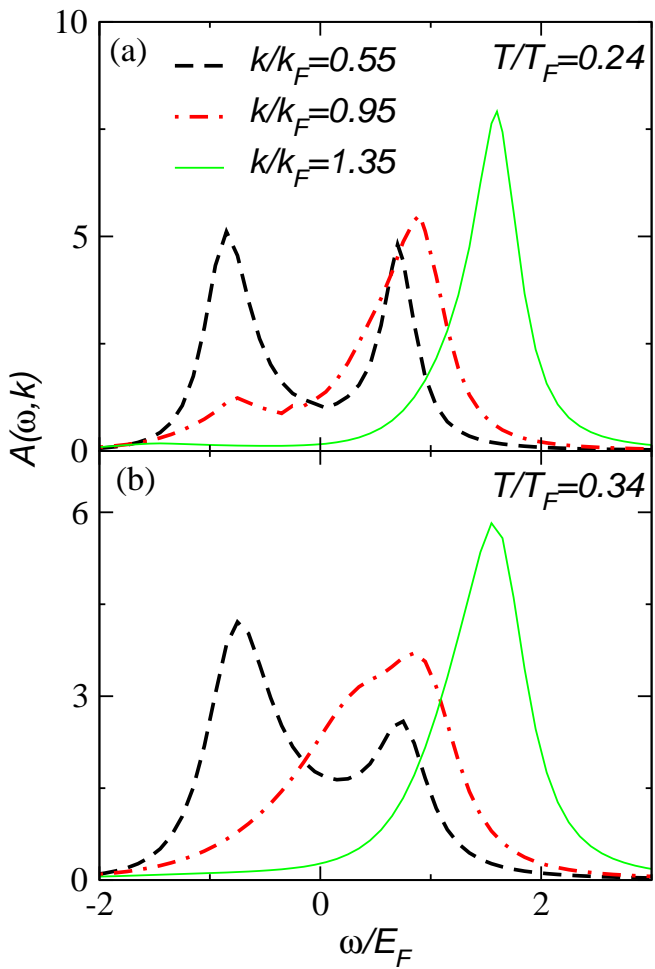}
\caption{Behavior of the frequency dependent spectral function at unitarity 
in NSR case for (a) $T/T_F = 0.24$ and (b) $T/T_F = 0.34$ for various wave-vectors ${\bf k}$.
This Figure suggests that the two-peak structure near $k_F$ associated with
this crossover theory barely meets the (most restrictive) 
definition for the presence of a pseudogap near $T_c$. Away from $T_c$, the two-peak structure near $k_F$ is virtually not observable.}
\label{fig:4g}
\end{figure}

\begin{figure} 
\includegraphics[width=3.4in,clip] 
{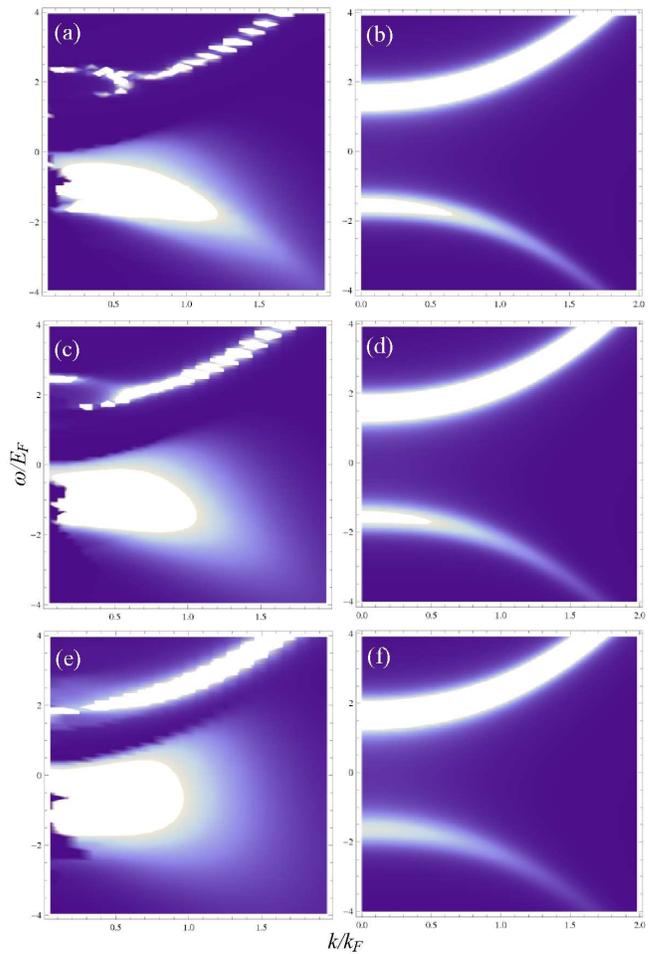}
\caption{(Color online) Spectral function obtained from NSR theory (left column) and from the extended BCS-Leggett theory (right column) at $1/k_{F}a=1$. Temperatures ($T/T_F$) from top to bottom are: (a) and (b) $0.25$, (c) and (d) $0.34$, (e) and (f) $0.55$. The ranges of $k/k_{F}$ and $\omega/E_F$ are $(0,2)$ and $(-4,4)$, respectively. }
\label{fig:Sp_Iap1} 
\end{figure} 

The issue of what constitutes the proper definition of a pseudogap is
an important one and we turn next to the first definition we introduced
above in which one requires that the spectral
function $A (\omega,{\bf k})$ as a function of $\omega$ exhibits two
peaks around $k \approx k_F$. 
It is clearly seen that in the BCS Leggett approach 
this definition is met for all three curves exhibited in
Figure~\ref{fig:Sp_Ia0}. Because it is more difficult to establish this 
for the NSR results plotted in Fig.~\ref{fig:Sp_Ia0}, in Fig.~\ref{fig:4g} 
we address this question more directly for two different temperatures
and a range of $k$ values near $k_F$.
It should be clear that at the lower temperature (which is slightly
above $T_c$), a pseudogap is seen in NSR theory, although we have
seen 
that the peak dispersion is not well described by
Eq.~(\ref{eq:3a}). This pseudogap should not be viewed as a broadened
BCS like feature.
At the higher temperature shown by the bottom panel of Fig.~\ref{fig:4g} there appears
to be no indication of a pseudogap according to the first definition.
Only a single peak is found in the spectral function near $k_F$.

We next explore analogous curves in the BEC regime and thereby
investigate how many-body bound states affect the distribution of weight 
in the 
spectral function. Figure~\ref{fig:Sp_Iap1} illustrates the spectral
function on the BEC side of resonance with $1/k_{F}a=1$ obtained from NSR theory
(left column) and from the extended BCS-Leggett theory (right column) at
selected temperatures. Here $T_c/T_F=0.22$ for the NSR case and
$T_c/T_F=0.21$ in the BCS-Leggett scheme. There is noise at
small $k$ in the spectral function of NSR theory which is presumably a 
numerical
artifact. In the presence of bound states
(in a many-body sense), the lower branch of the spectral function of
NSR theory shows a downward bending near $T_c$, but it can be
seen that this behavior
rapidly evolves
to
an upward dispersion as $T$ increases (Fig.~\ref{fig:Sp_Iap1}(e)). In
contrast, the spectral function from the extended BCS-Leggett theory exhibits
the downward dispersion at all $T$ indicated, rather similar to
the behavior in the superfluid phase.

One can further
see from the figure that in the extended BCS-Leggett theory the lower branch has
a much weaker spectral weight compared to that of the upper branch. This
derives from the same phenomenon as in the 
BCS-Leggett ground state, where the coefficient $v^{2}_{k}$ becomes
negligibly small in the BEC regime.
The behavior of the NSR spectral function is rather different 
from its counterpart in BCS-Leggett
theory at all three temperatures. If one plots the spectral function at
$k=k_F$ as a function of $\omega$ two peaks are present with the
upper peak much sharper and narrower than the lower, as also
reported in
Ref.~\cite{OhashiNSR}.

\subsection{Comparison of Density of States}
\label{sec:DOS}
\begin{figure}
\includegraphics[width=3.2in,clip] 
{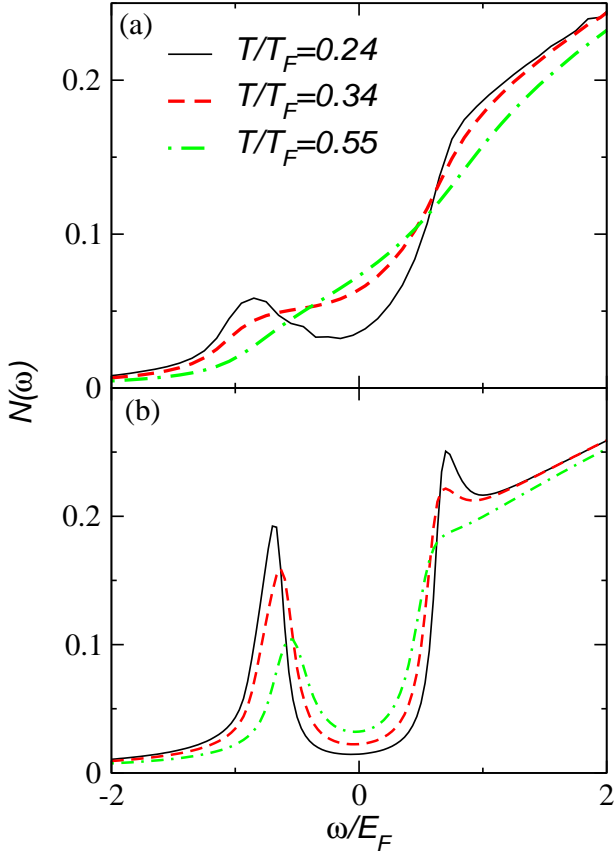}
\caption{(Color online) DOS at unitarity from (a) NSR theory and (b) the extended BCS-Leggett theory. (Black) solid line, (red) dashed line, and (green) dot-dash line correspond to $T/T_F=0.24$, $0.34$, and $0.55$.}
\label{fig:DOS}
\end{figure}

We turn now to the DOS which, when depleted around
the Fermi energy, provides a second
criterion
for the existence of the pseudogap.
The DOS is given by
\begin{equation}
N(\omega)=\sum_{\bf k}A(\omega,{\bf k}).
\end{equation}
In the HTSCs, an above-$T_c$- depletion in the DOS around the Fermi energy, 
measured in tunneling
experiments, provided
a clear
signature of the pseudogap (\cite{Timusk} and references therein).
By contrast, in the ultra-cold
Fermi gases the DOS has not been directly measured, although it is
useful to discuss it here in an abstract sense.

Figure~\ref{fig:DOS} shows the
DOS from the two theories at unitarity and 
at selected temperatures. 
The DOS based on the extended BCS-Leggett theory clearly shows a pseudogap
at all three selected temperatures. Similarly, the DOS from NSR theory
show a clear depletion around the Fermi energy ($\omega=0$) at $T/T_F=0.24$.
This depletion is barely visible at
$0.34$. At higher temperature ($T/T_F=0.55)$, the depletion does
not appear, and one sees only an
asymmetric background. We note that
our NSR
results are similar to those in Ref.~\cite{OhashiNSR} although
different number equations were employed. With this criterion for
a pseudogap one would conclude that 
NSR theory does have a pseudogap --- at least at $T\approx T_c$.
It is somewhat unlikely that the FLEX scheme (which at
$T_c$ seems to behave similarly to the highest $T$ NSR figures)
a pseudogap would be present--- via this second definition.

\subsection{Comparison of RF Spectra of Unpolarized Fermi Gases}
\label{sec:RF}
\begin{figure}
\includegraphics[width=3.2in,clip] 
{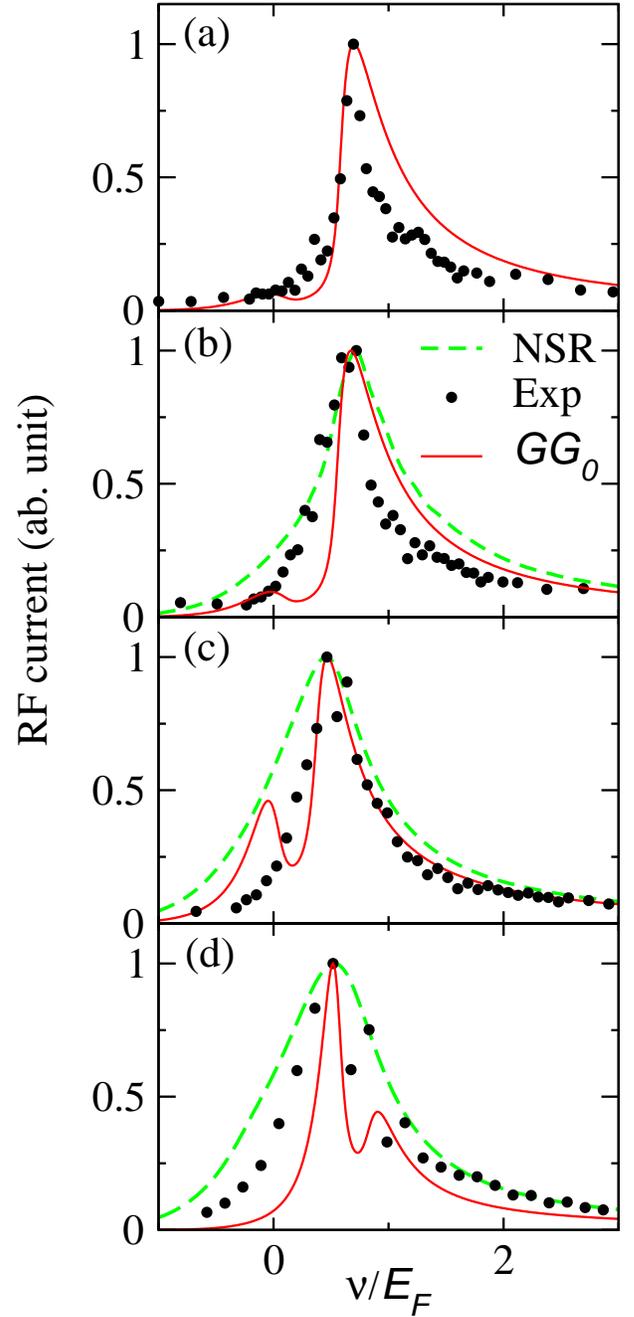}
\caption{(Color online) RF spectrum at $1/k_{F}a=0$. The (black) solid dots, (red) solid lines, and (green) dashed lines correspond to the RF currents obtained from the experimental data (\cite{MITtomo}, supplemental materials), the extended BCS-Leggett theory, and NSR theory. The values of $T/T_F$ are (a)$0.2$, (b)$0.22$ ($0.24$ for the curve from NSR theory), (c)$0.34$, and (d)$0.55$. }
\label{fig:RF_Ia0}
\end{figure} 
The RF current at detuning $\nu$ 
also depends on an integral involving the
fermionic spectral function. The current obtained from
linear response theory is given by \cite{heyan}
\begin{eqnarray}
I_0^{RF}(\nu)
&=& \sum_{\bf k} \left. \frac{|T_k|^2}{2\pi} A(\omega,{\bf k})
f(\omega)\right|_{\omega=\xi_{\bf k} -\nu}. 
\label{RFc0}
\end{eqnarray}
where, for the present purposes we ignore the complications 
from final-state effects \cite{Baym2,ourRF3}, as would be reasonable for the
so-called ``13" superfluid of $^{6}$Li. Here $|T_k|^2$ is a tunneling
matrix element which is taken to be a constant. The data points in
Figure~\ref{fig:RF_Ia0} correspond to measured tomographic spectra
from Ref.~\cite{MITtomo} in units of the local Fermi energy, (see the
Supplemental Materials). The results from NSR theory are indicated by
the dashed lines and from the extended BCS-Leggett theory ($GG_0$) by
the solid (red) lines.  To compare and contrast the spectra, we
normalize each curve by its maximum and align the maxima on the
horizontal axis, so as to effectively include Hartree shifts. The
experimental data were taken at $T/T_F=0.2, 0.22, 0.34, 0.55$. The RF
spectra from the extended BCS-Leggett theory are calculated at the same set of
temperatures. The RF spectra from NSR theory, which are restricted to
the normal phase correspond to the three higher temperatures:
$T/T_F=0.24, 0.34, 0.55$.

The RF spectra from the extended BCS-Leggett theory at high
temperatures indicate a double-peak structure, which was addressed in
Ref.~\cite{ourRF3}.
This peak at negative RF detuning emerges at finite temperatures in
BCS-Leggett theory as a result of thermally excited
quasiparticles. With increasing $T$, the weight under this peak
increases although the peak-to-peak separation will decrease,
following the temperature dependent pairing gap, as seen in the
figure. When temperature increases, the peak at negative RF detuning
grows and nearly merges with the peak at positive RF detuning so that
it may not be resolved experimentally.

By contrast, the RF spectra in NSR theory show a single peak which is
broader than the experimental RF spectra. This is to be expected based
on our analysis of the NSR spectral function in the previous section,
where we saw that the symmetrical upward and downward dispersing
branches of BCS theory were not present.  The RF spectra presented in
Ref.~\cite{HaussmannRF} using $GG$ t-matrix (FLEX) theory also shows a
broader (than experiment) single peak.

In view of the contrast between the BCS-Leggett curves and experiment,
it 
is natural to ask why is there no indication of the
negative detuning peak in these unpolarized experiments?
One can contemplate whether this stems from the
fact that (owing to large $\gamma$)
the two peaks simply aren't resolved. This would yield a figure
closer to that obtained from NSR-based calculations, which
is associated with a rather broad peak structure. As a result
it would not lead to a more satisfying fit to experiment.
At this stage we have no clear answer, but it will be important
to investigate, as we do below,
very slightly polarized gases to gain some insight.

\begin{figure}
\includegraphics[width=3.2in,clip] 
{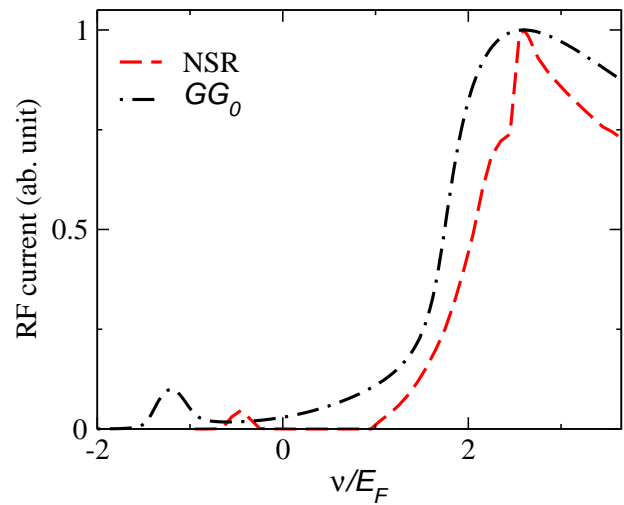}
\caption{(Color online) RF spectrum at $1/k_{F}a=1$. The (black) dot-dash line and (red) dashed line correspond to the RF currents obtained from the extended BCS-Leggett ($GG_0$) theory and NSR theory at $T/T_F=0.34$.}
\label{fig:RF_Iap1}
\end{figure}

In Figure~\ref{fig:RF_Iap1} we present the comparison on
the BEC side of resonance.
Here $1/k_{F}a=1$
and $T/T_F=0.34$. In this case both spectra show a double-peak
structure. For convenience,
we have scaled both spectra to their maxima and aligned the
maxima. For the NSR case,
the double-peaked feature reflects the negative fermionic 
chemical potential which is associated with
bound states. It similarly reflects the
stronger spectral weight of the upper branch in the spectral
function which can also be examined in Figure~\ref{fig:Sp_Iap1}. Our RF spectra in NSR theory are
consistent with
those presented in Ref.~\cite{StoofRF}.

\section{Extended BCS-Leggett Theory: Signature of $T_c$ in RF Spectrum}
\label{sec:RFTc}
\begin{figure}
\includegraphics[width=3.2in,clip] 
{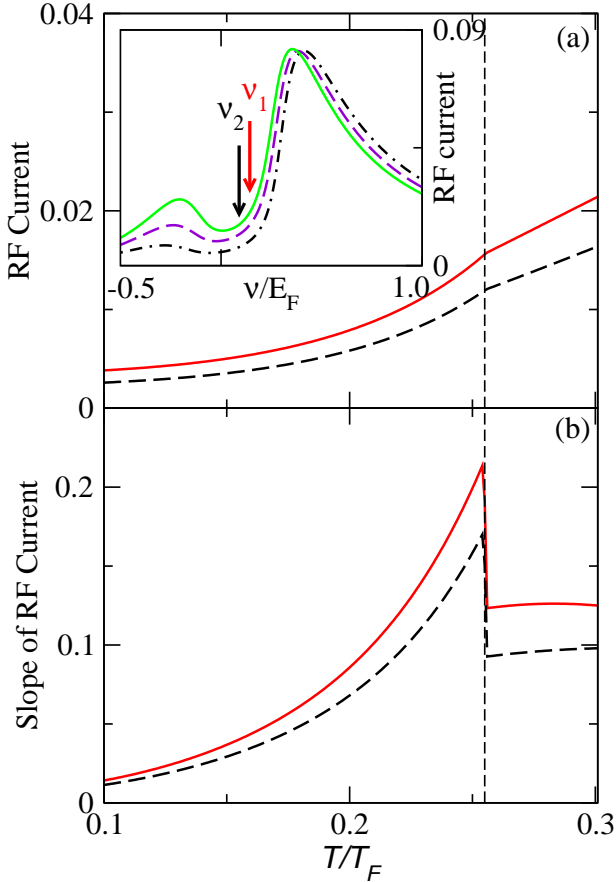}
\caption{(Color online) (a) RF current as a function of $T$ at $1/k_{F}a=0$ for $\nu_1/E_F=0.15$ (black dashed line) and $\nu_1/E_F=0.1$ (red solid line) obtained from the extended BCS-Leggett theory. Inset: RF currents as a function of detuning at $T/T_F=0.22$ (dot-dash line), $0.26$ (dashed line), $0.3$ (solid line). The two arrows indicate $\nu_1$ and $\nu_2$. (b) The slopes of the RF currents from (a). The vertical dashed line indicates $T_c$.}
\label{fig:RF_Tc_Ia0}
\end{figure} 

We have tried in the paper to emphasize comparisons whenever possible,
but there are instances where other crossover theories (besides that
based on the BCS-Leggett theory) have no counterpart.  In the first of these we
investigate the signature of the second order transition which should
be a subtle, but nevertheless thermodynamically required feature of
any crossover theory.  The experimental RF spectra in
Ref.~\cite{Grimm4,MITtomo} imply that the RF spectrum is more
sensitive to the existence of pairing rather than to
superfluidity. That it evolves smoothly across $T_c$ is due to the
presence of noncondensed pairs. The extended BCS-Leggett theory has the
important advantage in that it describes a smooth transition across
$T_c$ and should be a suitable theory for investigating this
question. In contrast, NSR theory and its generalization below $T_c$,
as well as the FLEX or Luttinger-Ward approach \cite{HaussmannRF}
encounter unphysical discontinuities

In the following we search for signatures of $T_c$ in the RF 
spectrum as obtained from BCS-Leggett theory near $T_c$. 
Here, in contrast to
Ref.~\cite{RFlong}, we use constraints provided by
our semi-quantitative fits to RF spectra (associated with
the estimated size of $\gamma(T)$) to obtain a more direct
assessment of how important these superfluid signatures should
be.

Figure~\ref{fig:RF_Tc_Ia0}
presents a plot suggesting how one
might expect to see signatures of coherence in a tomographic
(but momentum integrated) RF probe,
such as pioneered by the MIT group \cite{MITtomo}.
It shows the RF current versus temperature
at three different detuning frequencies. The inset plots the
RF characteristics indicating where the frequencies are chosen.
One can see that there is a feature at $T_c$, as expected.
This shows up more clearly in the lower figure which
plots the temperature derivative.
The same sort of feature has to be contained in the
specific heat \cite{ThermoScience}
which represents an integral over the spectral function.
What does it come from, since RF is not a phase sensitive
probe? The feature comes from the presence of a condensate
below $T_c$. What is distinguishing condensed from non-condensed
pairs is their self energy contribution. In the HTSCs \cite{Normanarcs}
and
also in the BCS-Leggett formulation the self energy from
the non-condensed pairs is taken to be
of a broadened BCS form in Eq.~(\ref{eq:9}).
By contrast, the non-condensed pairs live infinitely long and so have no
damping $\gamma$. These are the effects which are represented
in 
Figure~\ref{fig:RF_Tc_Ia0}.
In this way the
figure shows that there are features at $T_c$ which can in
principle help to distinguish the ordered state from the normal
pseudogap phase.

\section{Extended BCS-Leggett Theory: RF Spectrum of Polarized Fermi Gases}
\label{sec:RFpol}
\begin{figure}
\includegraphics[width=3.2in,clip] 
{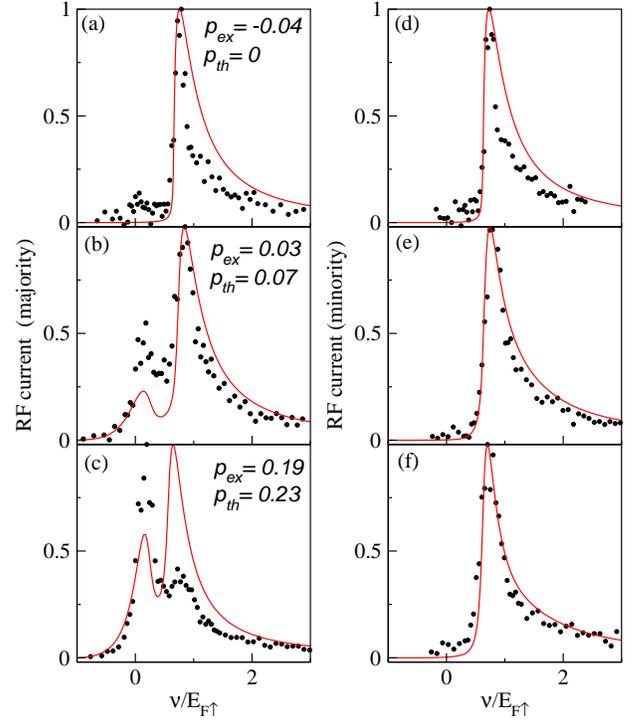}
\caption{(Color online) RF spectra of polarized Fermi gases at unitarity. (Black) dots and (red) solid lines correspond to the experimental data from Ref.~\cite{MITtomo} and results from the extended BCS-Leggett theory. The left (right) column shows the RF spectra for the majority (minority) species. The local temperature $T/T_{F\uparrow}$ and local polarization $p$ for the experimental data ($ex$) and the theoretical results ($th$), $(T_{ex}/T_{F\uparrow}, T_{th}/T_{F\uparrow}, p_{ex}, p_{th})$, are: $(0.05, 0.04, -0.04, 0)$ for (a) and (d); $(0.06, 0.13, 0.03, 0.07)$ for (b) and (e); $(0.06, 0.15, 0.19, 0.23)$ for (c) and (f).
}
\label{fig:RFImb_Ia0}
\end{figure} 

Another strength of the BCS-Leggett approach is that it can
address polarized gases at unitarity, which are not as readily
treated 
\cite{Parish,Hupolarized} in the alternative crossover theories.
In Figure~\ref{fig:RFImb_Ia0} we plot the RF spectra from the extended
BCS-Leggett theory and the experimental RF spectra from
Ref.~\cite{MITtomo}. Since the experimental RF spectra were obtained
from RF tomography of trapped polarized Fermi gases, we follow a
similar procedure to extract our RF spectra at varying, but comparable
locations from a similar trap profile. Also indicated are the
polarizations $p$. If we make fewer restrictions on the choice of
radial variable, the agreement is better as is shown in
Ref.~\cite{RFlong}. To compare the results, we normalized the maxima
and align the spectra, thereby introducing a fit to Hartree
contributions. The left (right) column shows the RF spectra of the
majority (minority). Here we set $\gamma/E_F=0.05$ and
$\Sigma_0/E_F=0.02$.

The experimental data points from the left hand
column can be compared with those in 
Figure 
~\ref{fig:RF_Ia0}
which are for the
$p \equiv 0$ case, and it is seen that even at very
small polarizations (say $p \approx 0.03$) the negative detuning
peak becomes visible.
Indeed, it appears here to be larger than the theoretically
estimated negative detuning peak height. 
A possible 
explanation for why the
double-peak structure can be resolved experimentally in polarized
but not in unpolarized Fermi gases is because the
existence of excess majority fermions causes a negative RF-detuning
peak even at low temperatures. At these lower $T$
the separation between the two
peaks can be large in the experimental RF spectra of polarized Fermi
gases. In contrast, for an unpolarized Fermi gas the negative
RF-detuning peak due to thermally excited quasiparticles only becomes
significant at high temperatures around $T_c$
and above. Here the separation between the two
peaks may not be as readily resolved. As expected, at low
temperatures there is only a single peak in the RF spectra of the
minority. We notice that at extremely high polarization, polaron-like
behavior has been observed in RF experiments \cite{Zwierleinpolaron},
whose explanation has attracted a great deal of attention
in the theoretical community \cite{Lobo,Chevy2}.
These effects have not been incorporated in our BCS-Leggett
formalism, where the normal state has been assumed to be
strictly non-interacting \cite{ChienPRL}.

\begin{figure}
\includegraphics[width=3.2in,clip] 
{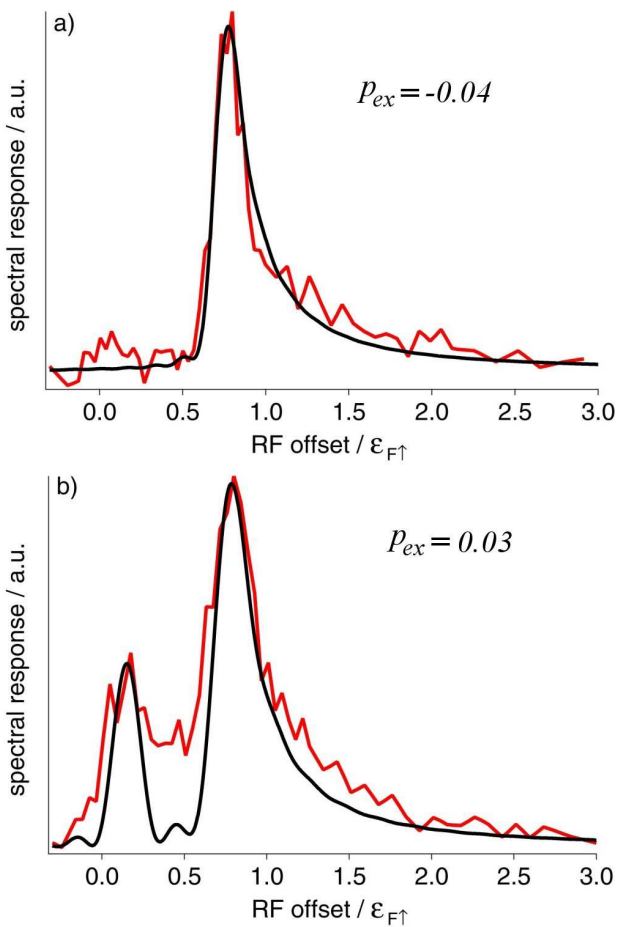}
\caption{(Color online) Reproduction of Fig.5 in the Supplemental Materials of Ref.~\cite{MITtomoimb}. Red curves correspond to experimental data from Fig.~\ref{fig:RFImb_Ia0} (a) and (b). Black curves are RF currents calculated from BCS-Leggett theory.
}
\label{fig:MITBCSfit}
\end{figure} 

The Ketterle group \cite{MITtomoimb} has argued that it should be
possible 
to extract the pairing gap size from RF spectroscopy in 
polarized gases at very low temperatures.
In Fig.~\ref{fig:MITBCSfit} we present a plot from their
paper (Supplementary Materials) which relates to their procedure.
This figure presents a fit to a generalized BCS-Leggett ground
state in the presence of polarization. The red curves correspond to
the actual data and the black curves are obtained from this
theory. An additional resolution broadening is included in the
theory and one can see that this theoretical approach appears
to be in quite reasonable agreement with experiment.
In this way there is some support for this simplest of ground
states --- at least in the polarized case.

\section{Conclusion}
\label{sec:conclusion}

The goal of this paper is to communicate that BCS-BEC crossover
theories are very exciting. They are currently being clarified and
developed hand in hand with experiment. For the Fermi superfluids,
unlike their Bose counterparts, we have no ready-made theory. In this
paper we confine our attention to the normal phase, although we have
presented a discussion of some of the controversial issues which have
surface in the literature below $T_c$.  We view the principal value of
this paper is the presentation of comparisons of two different
crossover theories and the identification of (mostly future)
experiments which can help distinguish them.  The two theories we
consider are the extended BCS-Leggett theory and that of Nozieres and
Schmitt-Rink.  We chose not to discuss the FLEX or Luttinger-Ward
scheme in any detail because it is discussed elsewhere
\cite{HaussmannRF}, and because there are concerns that, by ignoring
vertex corrections, this approach has omitted the important physics
associated with the pseudogap. These concerns are longstanding
\cite{Moukouri,Tremblay2,Fujimoto,Micnas95}.
Here we have argued that the extended BCS-Leggett theory is the one
theory which preserves (broadened) BCS features into the normal state
over a significant range of temperatures.  Even above $T_c$ one finds
that the fermionic excitations have an (albeit, smeared out)
dispersion of the form $E_{\bf k} \approx \pm \sqrt{ (\epsilon_{\bf k}
- \mu) ^2 + \Delta_{pg}^2}$ in the normal state.
 We find that
NSR theory does not have this dispersion, although it has a pseudogap
by all other measures. Interestingly high $T_c$ superconductors have
been shown to have this dispersion in their normal state \cite{ANLPRL}
and it is generally believed
\cite{Norman98,Normanarcs,FermiArcs}
that their fermionic self energy can be fit to a broadened ($d$-wave)
BCS form $\Sigma_{pg}(K) \approx
\Delta_{pg}({\bf k})^2/(\omega+\epsilon_{\bf k}-\mu+i\gamma)$.

In this paper we show that one can identify both physically and
mathematically the difference between the two normal states of the
different crossover theories. Mathematically because BCS theory
involves one dressed Green's function in the pair susceptibility, it
leads to a low frequency gap in the t-matrix or pair propagator (at
low $q$). Physically this serves to stabilize low momentum pairs. This
helps us to understand that the pseudogap of NSR theory does not
incorporate primarily low momentum pairs, but rather pairs of all
momentum and that it should be better further from condensation.
Indeed, this is reenforced by our observation that at higher $T$,
feedback effects which distinguish the two theories becomes less and
less important and the BCS-Leggett pair susceptibility, $GG_0$,
crosses over to something closer to $G_0G_0$ as in NSR theory.  Our
simplest approximation for the self energy in Eq.~(\ref{eq:2a}) is no
longer suitable once temperature exceeds, say $T/T_c \approx
1.5$. Indeed this is reenforced by earlier numerical observations
\cite{Maly1,Maly2}.

As a result, we believe that both theories are right but in different temperature
regimes.
Moreover, this serves to elucidate another concern about NSR theory
(and FLEX theory)--  that they are associated with an unphysical first order
transition. Both theories change discontinuously in going from
above to below $T_c$. In the superfluid phase the coupling which
is included in all other theories is between the non-condensed pairs
and the collective modes of the condensate, even though in the
normal state one couples the fermions and the non-condensed
pairs. In the extended BCS-Leggett theory, (as seems reasonable,
in the vicinity of $T_c$ both above and below), the dominant
coupling is, indeed,
between non-condensed pairs and fermions. These
(effectively, pseudogap) effects 
will behave smoothly across $T_c$. The Goldstone modes which turn on
at $T_c$ are 
highly damped in its vicinity, where the condensate is weak.  Only
at lower $T$ should their coupling become the more important.

In summary, a central conclusion of this study of the spectral
functions of the extended BCS-Leggett theory and NSR theory is that
one may expect that the former is suitable near $T_c$ due to its
similarity to BCS theory while the latter better describes the normal
phase at much higher $T$ as the system approaches a Fermi liquid, and
concomitantly, 
the pseudogap begins to disappear. In the course of this work we
have found that the theoretical RF spectra from both theories agree
(only semi-quantitatively) to about the same extent with experimental
data at unitarity. The BCS-Leggett approach has the advantage that it
can address the RF spectrum of generally polarized Fermi gases without
the problems which have been noted \cite{Hupolarized} for the NSR
approach.  However, momentum resolved experiments \cite{Jin6} may be
the ultimate way of distinguishing experimentally between different
theories.

\section*{Acknowledgement}
This work was supported by Grant Nos. NSF PHY-0555325 and NSF-MRSEC
DMR-0213745. We thank Prof. E.J.Mueller for helping substantially with the numerical calculations of NSR theory and Prof. Q.J.Chen for useful discussions.

\vspace*{-1ex} 

\bibliographystyle{apsrev}


\end{document}